\documentclass[preprint,12pt]{elsarticle}

\usepackage{amssymb}
\usepackage{color}

\journal{Physics Letters B}

\begin{document}

\begin{frontmatter}



\title{Constraining dark matter-nucleon scattering cross section by the background electron anti-neutrino flux data}


\author{Man Ho Chan, Chak Man Lee}

\address{Department of Science and Environmental Studies, The Education University of Hong Kong, Tai Po, New Territories, Hong Kong, China}

\ead{chanmh@eduhk.hk}

\begin{abstract}
Celestial objects such as stars and planets might be able to capture a large amount of dark matter particles through dark matter-nucleon scattering. Many previous studies have considered different celestial objects such as the Sun and the Earth as natural dark matter detectors and obtained some stringent bounds of the dark matter-nucleon scattering cross section. In this study, we use the $\sim 10$ MeV electron neutrino flux limits obtained by the Super-Kamiokande experiment and consider the Earth as a large natural dark matter detector to constrain the dark matter-nucleon scattering cross section. We show that this method can generally get more stringent limits. For certain ranges of dark matter mass annihilating via the $b\bar{b}$ channel, the limits of cross section for the isospin-independent scattering and proton-only scattering could be more stringent than that obtained in the PICO direct-detection experiment.
\end{abstract}


\end{frontmatter}



\section{Introduction}
The almost flat rotation curves in galaxies and unexpected large velocity dispersion in galaxy clusters reveal the existence of dark matter (DM). The standard cold dark matter (CDM) model suggests that the interaction (except gravity) between the particle DM and normal baryonic matter is extremely small. Nevertheless, such a very small interaction might still be detected based on current technologies. For example, the experiments PICO \cite{Amole,Amole2}, LUX \cite{Akerib}, XENON1T \cite{Aprile} and DEAP-3600 \cite{Yaguna} are going to detect DM signal directly via DM-nucleon interaction. However, no promising signals have been verified so far \cite{Aprile}. 

Besides the direct-detection experiments, it is also possible to constrain DM properties using celestial objects. The large and massive celestial objects such as the Sun \cite{Baum,Leane}, the Earth \cite{Baum,Mack,Bramante}, the neighbouring planets \cite{Bramante,Leane2}, white dwarfs \cite{McCullough} and neutron stars \cite{Leane3} can capture a large amount of DM particles and trap the DM inside the celestial objects. The captured DM particles would accumulate and finally annihilate to give electrons, positrons, neutrinos and gamma rays. The electrons, positrons and gamma rays would heat up the celestial objects while most neutrinos could leave the celestial objects completely. The bounds of the heat rate produced or neutrino flux detected can be used to constrain the cross section between DM and baryonic matter. For example, previous studies have examined the heat rate produced due to DM annihilation inside the Earth \cite{Mack,Bramante}, the Moon \cite{Garani2}, the Mars \cite{Bramante} and the Jupiter \cite{Leane2} to constrain the cross section. Some other studies have examined the neutrino flux produced in the Sun \cite{Baum}, the Earth \cite{Baum} and the Moon \cite{Chan} to constrain the cross section. Many of the limits are comparable to that of the direct-detection experiments. Therefore, using celestial objects as detectors can give complementary constraints for comparison.

In this study, we use the Earth as a natural DM detector and use the electron anti-neutrino flux measured to constrain the DM-nucleon cross section. Previous studies have considered the neutrino fluxes of energy GeV or above to calculate the constraints \cite{Baum}. Here, we use the anti-electron neutrino flux limit of neutrino energy $\sim 10$ MeV or above to do the analysis. We show that using the $\sim 10$ MeV neutrino flux can give generally more stringent limits of the DM-nucleon cross section compared with the other studies using celestial objects as detectors. 

\section{The DM-nucleus scattering model}
Due to the relative motion between the DM particles and the Earth, some DM particles would probably scatter off the atomic nuclei and electrons of the Earth. During the scattering, the DM particles would lose energy so that their resulting velocities are lower than the Earth's escape velocity $v_e$, and will finally be trapped inside the Earth. Although the interaction rate between DM and atomic nuclei is very small, the large number of target nuclei in the Earth can at least capture a certain amount of DM particles inside the Earth. The capture rate $C$ depends on the Earth composition. We follow the benchmark model of the Earth composition to calculate the capture rate (see Table 1 for the number percentages of elements in different laye\emph{}rs including the crust, mantle and core) \cite{Bramante}. The capture rate of the DM particles of mass $m_{\chi}$ with the scattering cross section $(\sigma_{\chi i})$ of DM on collision nuclei of type $i$ with number $n_i$ is then given by \cite{Gould1, Gould2}
\begin{eqnarray}
C&=&\sqrt{\frac6{\pi}}\frac{\rho_{\rm DM}}{m_{\chi}v_d}
\sum_{i}\frac{\sigma_{\chi i} n_i}{2\eta A_i^2}\nonumber\\
&&\times\left[\left(A_{i,+}A_{i,-}-\frac12\right)(\chi(-\eta,\eta)-\chi(A_{i,-},A_{i,+}))\right.\nonumber\\
&&\left.+\frac12A_{i,+}\exp(-A_{i,-}^2)-\frac12A_{i,-}\exp(-A_{i,+}^2)\right.\nonumber\\
&&\left. -\eta\exp(-\eta^2)\right]
,\label{capture}\\
{\rm with}&&\nonumber\\
A_i^2&=&\frac{6v_e^2}{v_d^2}\frac{m_{\chi}m_i}{(m_{\chi}-m_i)^2},\;\;\;\;\;\;\; {\rm and }\,\, A_{i,\pm}=A_i\pm\eta,
\end{eqnarray}
where the dimensionless quantities $\chi(a,b)=(\sqrt{\pi}/2)[{\rm Erf}(b)-{\rm Erf}(a)]$ and $\eta=3u^2/2v_d^2$ with the velocity of Earth $u = 220~{\rm kms^{-1}}$. $\rho_{DM} = 0.3~{\rm GeV cm^{-3}}$ is the local DM density and the escape velocity is $v_e = 11.2~{\rm kms^{-1}}$. Here the Earth's motion with respect to the DM distribution is neglected for simplicity, and the latter is assumed to be Maxwellian with the velocity dispersion $v_d = 270~{\rm kms^{-1}}$. We sum up the contributions of different elemental nuclei in all of the three layers (crust, mantle and core). The factor in Eq.~(2) is responsible for a composition-dependent resonance-like behavior of the capture rate, which gives sharp peaks each corresponding to a particular elemental nuclei in the Earth.

In the followings, we will consider both spin-dependent and spin-independent scatterings between the DM and baryonic nuclei. Assuming isospin equivalence, for a nucleus with atomic mass $\bar{A}_i$, the DM-nucleus spin-independent elastic scattering cross section in Eq.~(\ref{capture}) is expressed, in terms of the DM-nucleon cross section $\sigma_{\chi N}^{\rm(SI)}$, by
\begin{equation}
\sigma_{\chi i}^{\rm (SI)}=\left(\frac{\mu_r^{\bar{A}_i}}{\mu_r^{N}}\right)^2\bar{A}_i^2\sigma_{\chi N}^{\rm (SI)},
\end{equation}
where $\mu_r^N=m_{\chi}m_N/(m_{\chi}+m_N)$ and $\mu_r^{\bar{A}_i
}=m_{\chi}m_{\bar{A}_i}/(m_{\chi}+m_{\bar{A}_i})$ are the reduced masses of proton-DM and nucleus-DM, respectively.

For the spin-dependent scattering, the cross section is parameterized by three nuclear values including the nuclear spin of an atom $J_i$, its average proton and neutron spins($\langle S_p\rangle$ and $\langle S_n\rangle$), and proton and neutron coupling constants ($a_p$ and $a_n$). The spin-dependent cross section can be given by \cite{Bramante}
\begin{equation}
\sigma_{\chi i}^{\rm (SD)}=\left(\frac{\mu_r^{\bar{A}_i}}{\mu_r^{N}}\right)^2\frac{4(J_i+1)}{3J_i}\left[a_p\langle S_p\rangle_i+a_n\langle S_n\rangle_i\right]^2\sigma_{\chi N}^{\rm (SD)}.
\end{equation}
The number percentages of elements with non-zero nuclear spin and their spin parameters are listed in Table 2 \cite{Bramante,Bednyakov}. We will consider three types of scattering: 1. isospin-independent scattering $(a_p=a_n=1)$, 2. proton-only scattering $(a_p=1, a_n=0)$, and 3. neutron-only scattering $(a_p=0, a_n=1)$. In Fig.~1, we show the capture rate $C$ by assuming a standard value of $\sigma_{\chi N}^{\rm(SI)}$ (or $\sigma_{\chi N}^{\rm(SD)}$) $= 10^{-35}{\rm cm}^{2}$. Generally speaking, the capture rate for the spin-independent scattering is the largest.

If DM can self-annihilate, the captured DM particles inside the Earth would self-annihilate to give high-energy particles such as electrons, positrons, photons and neutrinos (including anti-neutrinos). The electrons, positrons and photons would be quickly scattered by the Earth nuclei and contribute to internal heat while the neutrinos produced would almost completely leave the Earth. Therefore, by measuring the number of neutrinos passing through the underground neutrino detectors, we can estimate the DM annihilation rate. The time evolution of the DM particles gravitationally captured by it is given by
\begin{equation}
\frac{{\rm d}N_{\chi}}{{\rm d}t}=C-AN_{\chi}^2,
\label{dNdt}
\end{equation}
where $AN_{\chi}^2$ governs the number of DM particles ($N_{\chi}$) lost due to their annihilation. Here, we did not involve the evaporation term in Eq.~(\ref{dNdt}) as many recent studies have shown that the evaporation of DM inside the Earth is important only for $m_{\chi} \le 12$ GeV \cite{Garani}. In our study, we will mainly focus on $m_{\chi} \ge 20$ GeV so that we can neglect the evaporation term. Generally speaking, the annihilation rate is a function of time \cite{Leane3}. Nevertheless, as the age of the Earth is long enough ($\sim 4.5$ billion years) so that an equilibrium state would be achieved (i.e. $C=AN_{\chi}^2$). In this case, the annihilation rate is approximately equal to half of the capture rate $(C/2)$ because two DM particles are involved in each annihilation event.

Assume that the DM distribution is spherically symmetric and the radius of Earth is sufficiently large, $R_E \approx 6371$ km. The neutrinos emitted inside the Earth due to DM annihilation as observed from the Earth's surface can be regarded as point-source emission and the neutrino flux can then be expressed as
 \begin{equation}
S_{\nu}=\frac{1}{4\pi R_E^2}\left(\frac{C}{2}\right)\int_{E_{0}}^{m_{\chi}}\frac{{\rm d}N_{\nu,{\rm inj}}(E,m_{\chi})}{{\rm d}E}dE,
\label{SDM}
\end{equation}
where ${{\rm d}N_{\nu,{\rm inj}}(E,m_{\chi})}/{{\rm d}E}$ is the injected energy spectrum of DM annihilation contributed by a particular type of neutrinos. In particular, the injected energy spectrum depends on the annihilation channels. We will consider four popular annihilation channels ($e^+e^-$, $\mu^+\mu^-$, $\tau^+\tau^-$ and $b\bar{b}$) to constrain the DM-nucleon cross section. The injected neutrino energy spectrum can be obtained in \cite{Cirelli}. Here, we do not consider the effect of neutrino oscillation for simplicity. The neutrino oscillation length depends on the energy of neutrinos and it is much smaller than the Earth's radius. On average about half of the electron neutrinos produced in DM annihilation would probably change to the other two types during diffusion. Also, a certain amount of muon neutrinos and tau neutrinos produced may also oscillate to electron neutrinos. Overall speaking, the effect of oscillation does not have a significant impact on the electron neutrino and anti-neutrino fluxes produced.

\section{Results}
We use the MeV electron anti-neutrino data collected in the Super-Kamiokande (SK) experiment to constrain the annihilation rate (i.e. the capture rate). Then, following the theoretical framework, we can deduce the upper limits of the spin-dependent and the spin-independent cross sections based on the constrained annihilation rate. The neutrino data collected in the SK experiment were originally used to constrain the supernova relic neutrinos (SRNs) \cite{Bays,Zhang}. These SRNs can be regarded as the background electron anti-neutrinos passing through the Earth. Therefore, the neutrino data can also be used to constrain the annihilation rate originated from the DM captured by the Earth. 

The upper limit of electron anti-neutrino flux obtained by the SK experiment is $\approx 2.9$ cm$^{-2}$ s$^{-1}$ for neutrino energy $E>17.3$ MeV \cite{Bays}. We therefore calculate the neutrino flux in Eq.~(\ref{SDM}) with $E_0=17.3$ MeV and set $S_{\nu} \le 2.9$ cm$^{-2}$ s$^{-1}$ to calculate the upper limits of $\sigma_{\chi N}^{\rm (SI)}$ and $\sigma_{\chi N}^{\rm (SD)}$. In Fig.~2, we show the upper limits of the spin-independent cross section for the four popular annihilation channels. The limit for the $e^+e^-$ channel are less stringent because this channel produces less electron neutrino pairs, while the $b\bar{b}$ channel can give the most stringent limit among the four population channels. Generally speaking, our limits are more stringent than most of the limits obtained by previous studies using celestial objects as detectors. In particular, our limit for the $b\bar{b}$ channel is very close to the DEAP-3600 direct-detection limit for $m_{\chi}=20-50$ GeV. 

In Figs.~3-5, we show the upper limits of the spin-dependent cross sections for the four popular annihilation channels (3 types of scattering). Our study can give very stringent constraints, particularly for the $b\bar{b}$ channel. The limits for the isospin-independent and proton-only scatterings are the most stringent for $m_{\chi}=20-40$ GeV compared with the limits obtained by using celestial objects as detectors. In particular, for certain small ranges of $m_{\chi}$ (at the resonant troughs), the limits can be even more stringent than the direct-detection limits obtained by the PICO experiments (PICO-60 and PICO-2L) \cite{Amole,Amole2} (see Fig.~3 and Fig.~4).

\begin{figure}
\vskip 3mm
\includegraphics[width=140mm]{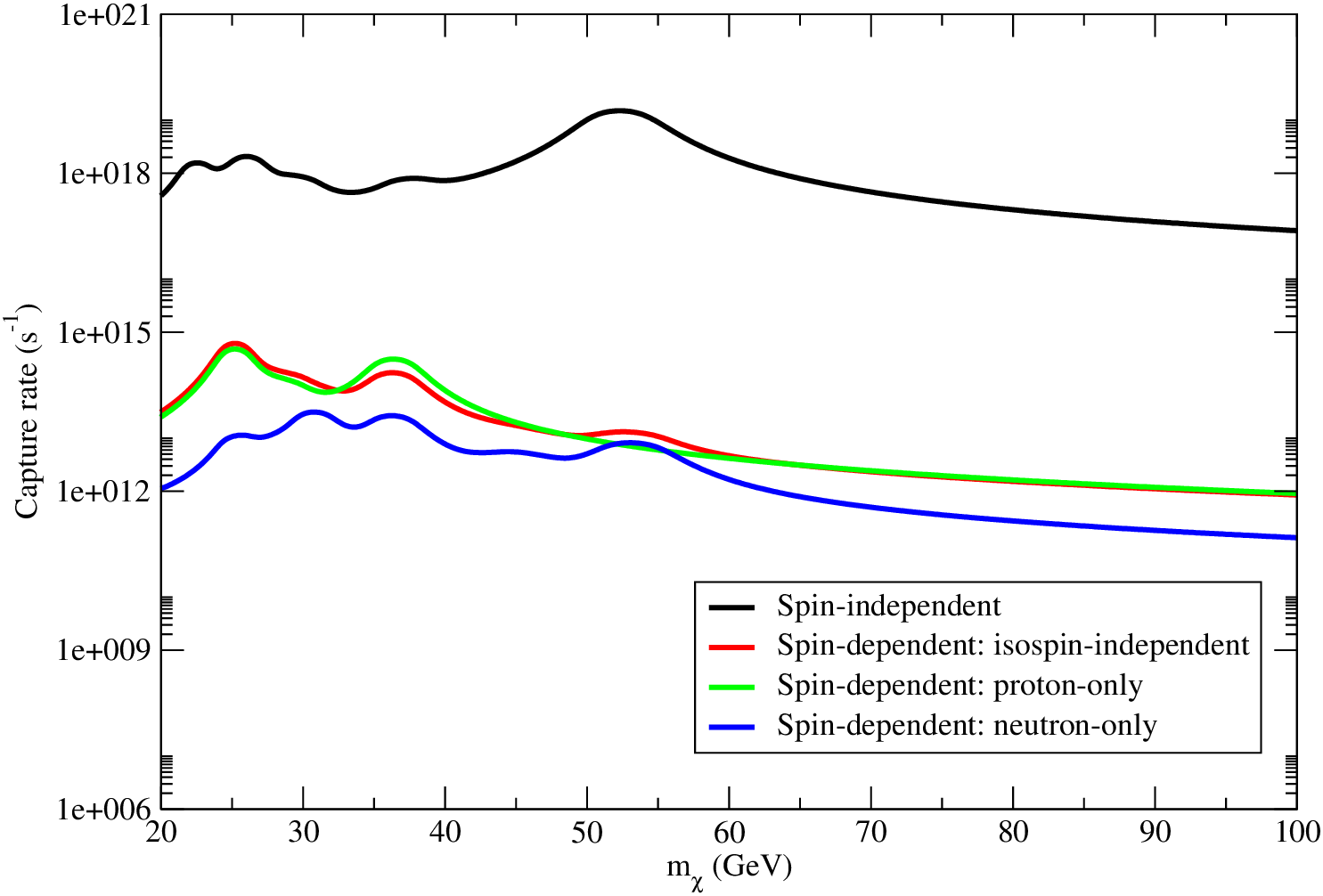}
\caption{Capture rate of dark matter for both spin-independent and spin-dependent cases, the latter case including isospin-independent scattering ($a_p = 1, a_n = 1$), proton-only scattering ($a_p = 1, a_n = 0$), and neutron-only scattering ($a_p = 0, a_n = 1$), when assumed $\sigma_{\chi N}^{\rm(SI)}(\sigma_{\chi N}^{\rm(SD)}) = 10^{-35}{\rm cm}^{2}$.}
\label{Fig1}
\vskip 3mm
\end{figure}

\begin{figure}
\vskip 3mm
\includegraphics[width=140mm]{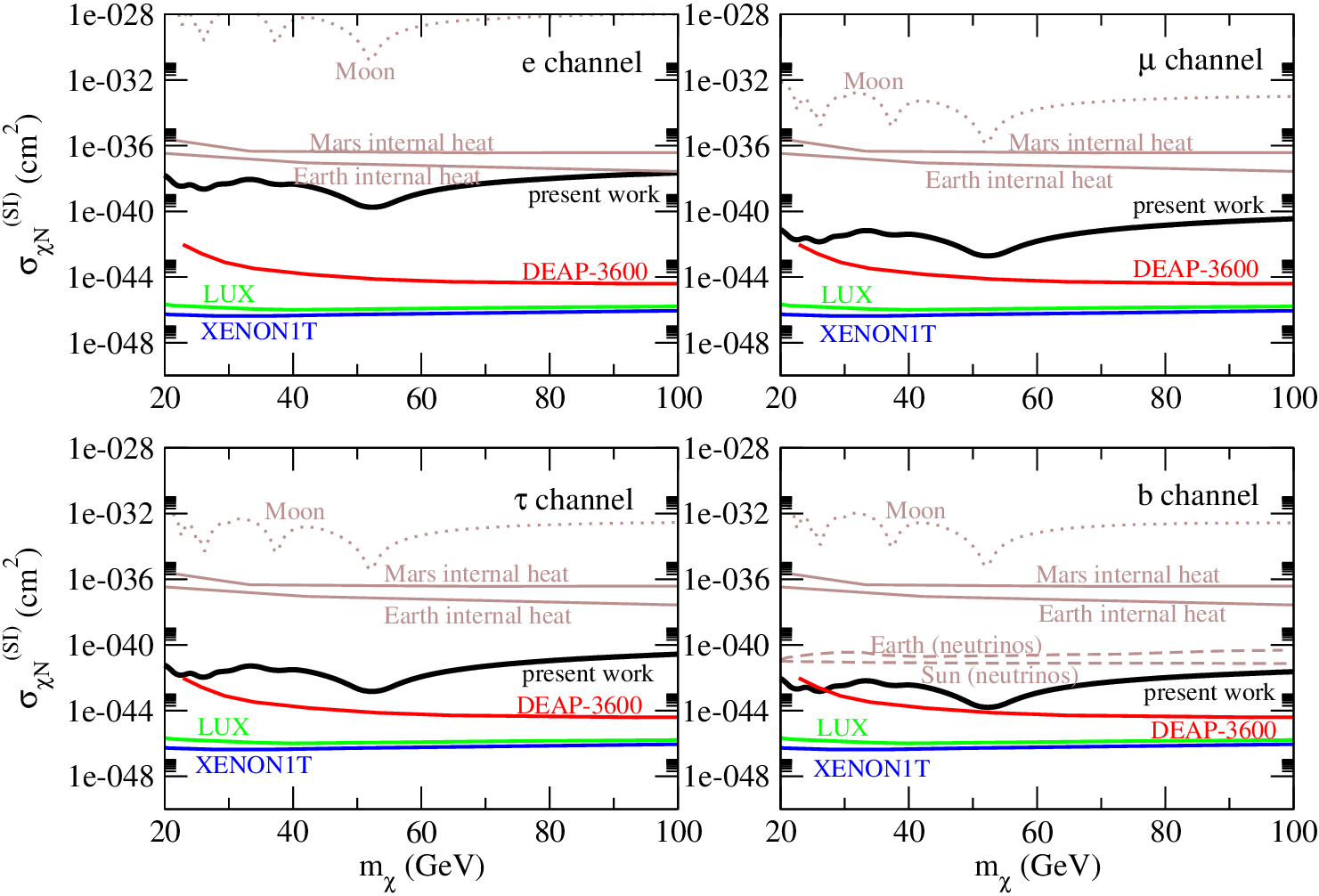}
\caption{The black solid lines represent the upper limits of $\sigma_{\chi N}^{\rm(SI)}$ for dark matter annihilation via $e^+e^-$ (top-left), $\mu^+\mu^-$ (top-right), $\tau^+\tau^-$ (bottom-left), and $b\bar{b}$ (bottom-right) channels for the spin-independent scenario. We also include the upper limits obtained from other studies using celestial objects as detectors (brown lines) \cite{Baum,Bramante,Chan} and the direct-detection limits (red, green and blue lines) \cite{Akerib,Aprile,Yaguna} for comparison.}
\label{Fig2}
\vskip 3mm
\end{figure}

\begin{figure}
\vskip 3mm
\includegraphics[width=140mm]{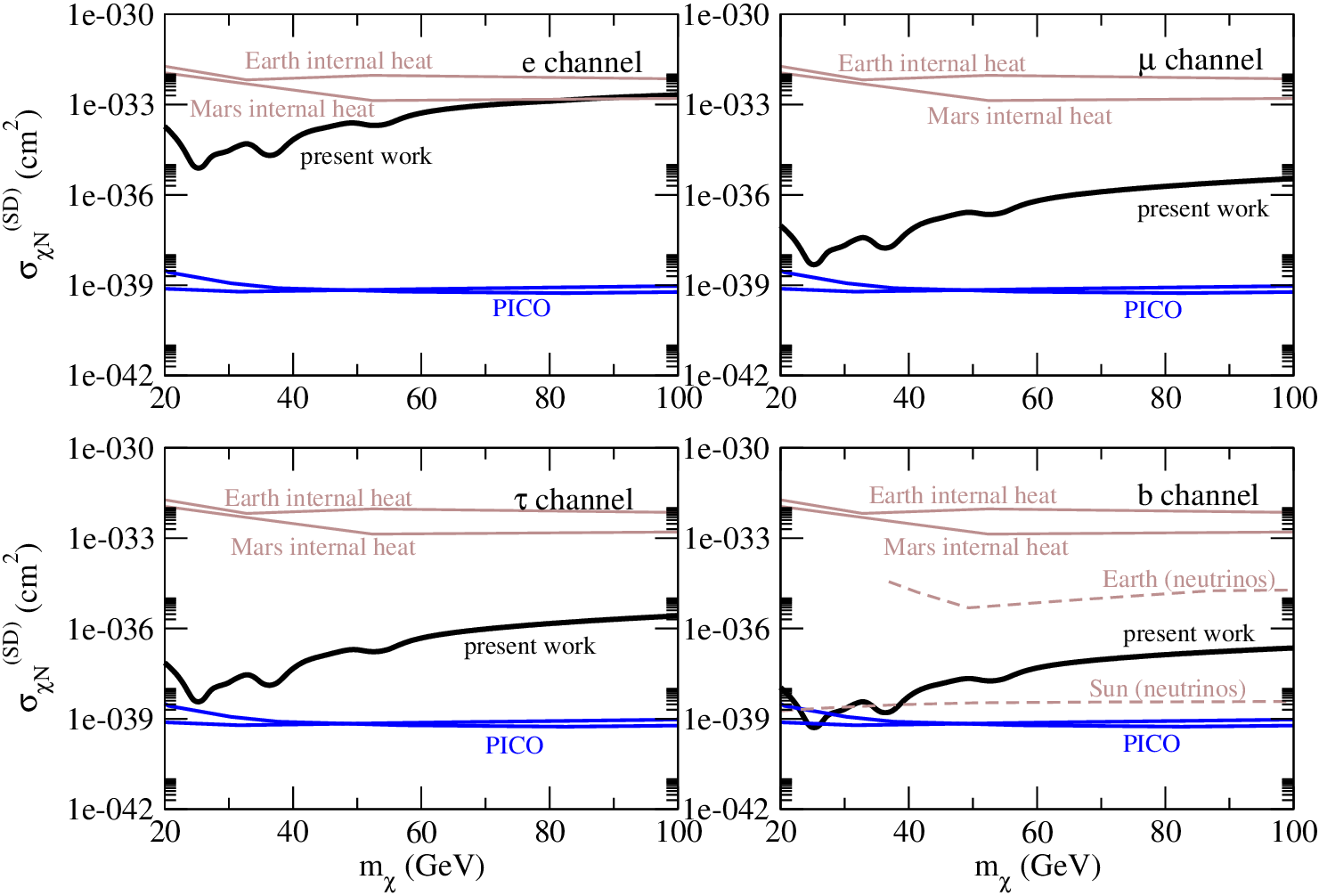}
\caption{The black solid lines represent the upper limits of $\sigma_{\chi N}^{\rm(SD)}$ for dark matter annihilation via $e^+e^-$ (top-left), $\mu^+\mu^-$ (top-right), $\tau^+\tau^-$ (bottom-left), and $b\bar{b}$ (bottom-right) channels for the spin-dependent scenario with $a_n=a_p=1$ (isospin-independent scattering). We also include the upper limits obtained from other studies using celestial objects as detectors (brown lines) \cite{Baum,Bramante} and the direct-detection limits of the PICO experiments (PICO-60 and PICO-2L) (blue lines) \cite{Amole,Amole2} for comparison.}
\label{Fig3}
\vskip 3mm
\end{figure}

\begin{figure}
\vskip 3mm
\includegraphics[width=140mm]{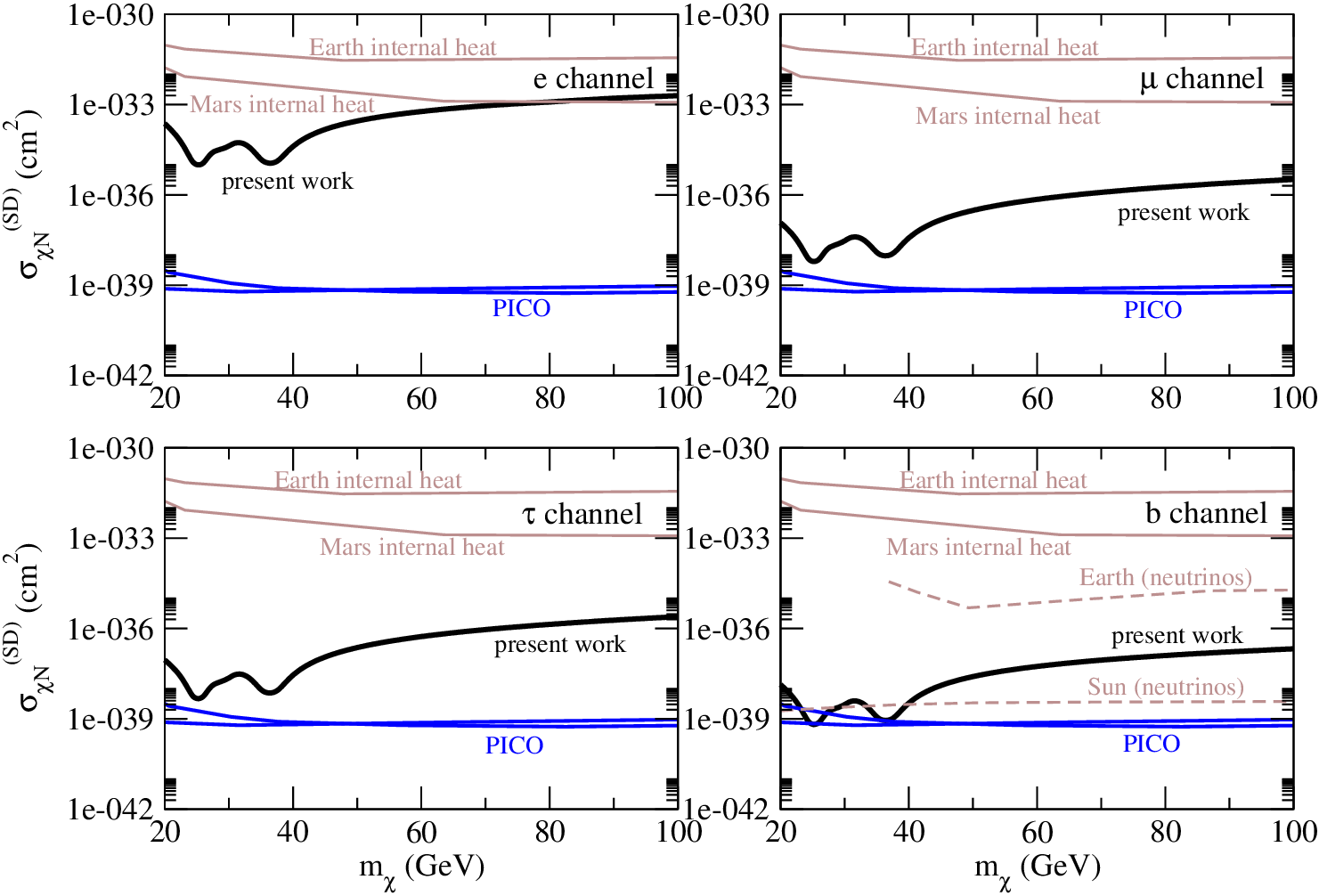}
\caption{Same as Fig.~3, for the case of proton-only scattering ($a_n = 0, a_p =1$).}
\label{Fig4}
\vskip 3mm
\end{figure}

\begin{figure}
\vskip 3mm
\includegraphics[width=140mm]{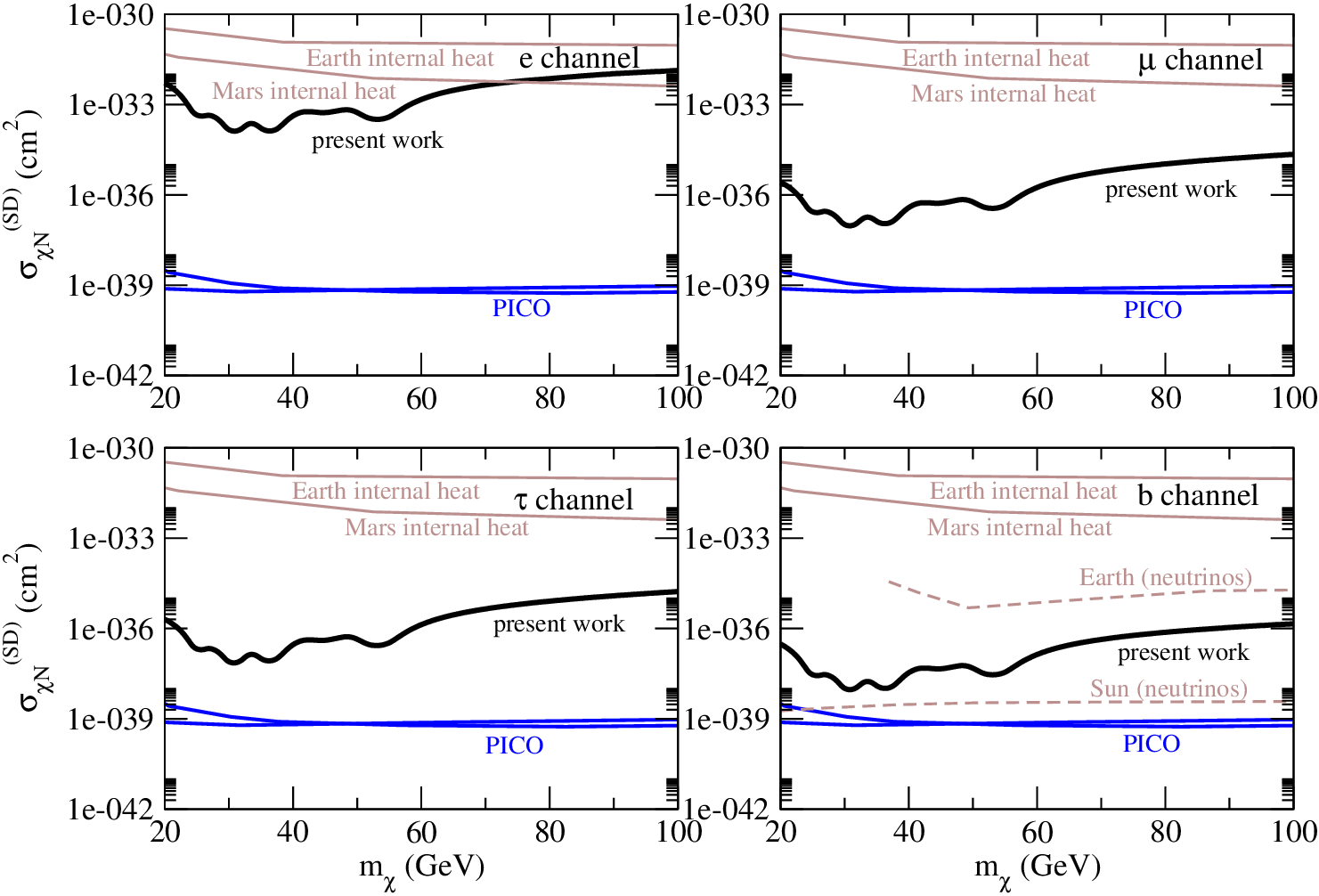}
\caption{Same as Fig.~3, for the case of neutron-only scattering ($a_n = 1, a_p = 0$).}
\label{Fig5}
\vskip 3mm
\end{figure}

\begin{center}
\begin{table}
\caption{The elemental abundances of the Earth \cite{Bramante}.}
\begin{tabular}{ |c|c|c|c| }
 \hline\hline

          & Radius (km)          & Radius (km)          & Radius (km)\\
          & 0 - 3480             &   3480 - 6346        &6346 - 6371\\
 \hline
&  Core                & Mantle               & Crust \\
Element   &no. of nuclei  &no. of nuclei   &no. of nuclei  \\
          &[$\times10^{48}$]      &[$\times10^{48}$]      &[$\times10^{48}$]                  \\
  \hline
$^{16}$O  &-                     &66.763                &0.376\\
$^{28}$Si &-                     &18.343                &0.127\\
$^{27}$Al &-                     &2.054                 &0.039\\
$^{56}$Fe &17.687                &2.713                 &0.012\\
$^{40}$Ca &-                     &1.507                 &0.012\\
$^{23}$Na &-                     &-                     &0.016\\
$^{39}$K  &-                     &-                     &0.009\\
$^{24}$Mg &-                     &22.405                &0.011\\
$^{48}$Ti &-                     &-                     &0.002 \\
$^{57}$Ni &1.152                 &0.085                 &-\\
$^{59}$Co &0.060                 &-                     &-\\
$^{31}$P &0.227                  &-                     &-\\
$^{32}$S &3.297                  &-                     &-\\
  \hline \hline
\end{tabular}
\end{table}
\end{center}

\begin{center}
\begin{table}
\caption{Number percentages of elements with non-zero nuclear spin in the Earth and their spin parameters \cite{Bramante,Bednyakov}}
\begin{tabular}{ |c|c|c|c|c| }
 \hline\hline
Element     &Number $\%$    &J            &$\langle S_p\rangle $    &$\langle S_n\rangle $\\
  \hline
$^{17}$O    &0.4            &5/2          &-0.036                   &0.508\\
$^{29}$Si   &4.7            &1/2          &0.054                    &0.204\\
$^{27}$Al   &100            &5/2          &0.333                    &0.043\\
$^{57}$Fe   &2.12           &1/2          &0                        &0.5\\
$^{43}$Ca   &0.135          &7/2          &0                        &0.5\\
$^{23}$Na   &100            &3/2          &0.2477                   &0.0199\\
$^{39}$K    &100            &3/2          &-0.196                   &0.055\\
$^{25}$Mg   &10             &5/2          &0.04                     &0.376\\
$^{47}$Ti   &7.44           &5/2          &0                        &0.21\\
$^{49}$Ti   &5.41           &7/2          &0                        &0.29\\
$^{61}$Ni   &1.14           &3/2          &0                        &-0.357\\
$^{31}$P    &100            &1/2          &0.181                    &0.032\\
$^{33}$S    &0.75           &3/2          &0                        &-0.3\\
  \hline \hline
\end{tabular}
\end{table}
\end{center}

\section{Discussion}
In this article, we use the background $\sim 10$ MeV electron anti-neutrino data to constrain different types of the dark matter-nucleon scattering cross section. The dark matter is assumed to be captured by the Earth and then self-annihilate to give neutrinos. The upper limits of the electron anti-neutrino flux detected can be used to constrain the capture rate and the scattering cross section. By examining four popular annihilation channels, we can get very stringent limits for the scattering cross section. 

Although the limits are generally less stringent than that of the direct-detection experiments, they are almost the most stringent among the limits obtained by using celestial objects (e.g. Sun, Mars) as detectors. Also, for the $b\bar{b}$ channel, the limits for $m_{\chi}=20-50$ GeV are quite close to the direct-detection limits. In particular, the upper limits of the spin-dependent cross section for certain ranges of $m_{\chi}$ can be even more stringent than the PICO direct-detection limits. In fact, some recent studies of dark matter annihilation have shown that a particular range of dark matter mass $m_{\chi}=30-50$ GeV annihilating via the $b\bar{b}$ channel can account for the gamma-ray excess in our Galaxy \cite{Daylan}, and the radio excess in our Galaxy \cite{Chan2} and some galaxy clusters \cite{Chan3,Chan4}. Therefore, the stringent limits of the scattering cross section in this range of dark matter mass obtained in our study can provide important complementary information for revealing the nature of dark matter.

In this study, we have used the data of the background $\sim 10$ MeV electron anti-neutrino flux limit. If we have a more stringent upper limit of the background $\sim 10$ MeV electron anti-neutrino flux, we can have more stringent upper limits of the DM-nucleon cross section. Therefore, we anticipate that a future better constraint of the background neutrino flux could further constrain the scattering cross section, which might be more stringent than the upper limits obtained by the current direct-detection experiments. 

\section{Acknowledgements}
The work described in this paper was partially supported by the Seed Funding Grant (RG 68/2020-2021R) and the Dean's Research Fund of the Faculty of Liberal Arts and Social Sciences, The Education University of Hong Kong, Hong Kong Special Administrative Region, China (Project No.: FLASS/DRF 04628).





\end{document}